\documentstyle[12pt,epsfig]{article}

\def\beq{\begin{equation}}   \def\eeq{\end{equation}}
\def\lsim{\mathrel{\rlap{\lower3pt\hbox{\hskip0pt$\sim$}}
    \raise1pt\hbox{$<$}}}         
\def\gsim{\mathrel{\rlap{\lower4pt\hbox{\hskip1pt$\sim$}}
    \raise1pt\hbox{$>$}}}         
\begin{document}
\def\simlt{\mathrel{\raise.3ex\hbox{$<$\kern-.75em\lower1ex\hbox{$\sim$}}}}
\def\simgt{\mathrel{\raise.3ex\hbox{$>$\kern-.75em\lower1ex\hbox{$\sim$}}}}
\begin{flushright}
TPI--MINN--01/56\\
UMN--TH--2037/01
\end{flushright}

\begin{center}
\baselineskip25pt

\vspace{1cm}

{\Large\bf
Flavordynamics with Conformal Matter and   Gauge Theories on Compact
Hyperbolic Manifolds in Extra Dimensions}

\vspace{1cm}

{\bf
D. A. Demir and M.  Shifman}

\vspace{0.3cm}

Theoretical Physics Institute,
University of Minnesota, Minneapolis, MN 55455

\vspace{1cm}

Abstract

\end{center}

We outline a toy model in which  a unique mechanism may  trigger a dynamical chain
resulting in key low-energy
regularities. The starting points are a negative cosmological term in the bulk
and conformally invariant nongravity sector. These elements
ensure compactification of the extra dimensional space  
 on a compact hyperbolic manifold (with the negative and constant scalar curvature).
The overall geometry is then  ${\bf M}_{4}\times {\bf B}_{n}$.
The negative curvature on  ${\bf B}_{n}$ triggers the formation of the four--dimensional
defect which provides in turn a dynamical localization of  ordinary particles. 
It also leads, simultaneously, to a spontaneous breaking of gauge symmetry
through a Higgs mechanism. Masses of the fermions, gauge bosons and scalars 
all derive from the curvature of the internal manifold such that the Higgs 
boson is generally heavier than the gauge bosons. The factorizable geometry 
${\bf M}_{4}\times {\bf B}_{n}$ and flatness of ${\bf M}_{4}$ require fine-tuning.



\newpage


\section{Introduction}

Theories with large extra spatial dimensions
have allowed one to reformulate the hierarchy problem
in a geometric paradigm \cite{add,ant,add1}.
 If space-time is in fact $(4+n)$-dimensional
 the Planck scale of gravity in four dimensions, $M_{\rm Pl}$,
is determined by the fundamental  ($4+n$)-dimensional scale, $M_{\star}$, and
geometry   of the extra space. In the simplest case, when the
 space-time is a product of the four-dimensional
Minkowskian space-time  ${\bf M}_{4}$ and  an $n$-dimensional compact space
${\bf B}_{n}$, the two gravitational scales are related as follows:
\begin{equation}
M_{\rm {Pl}}^{2} = {\cal{V}}_{n} M_{\star}^{n+2}
\label{one}
\end{equation}
 where ${\cal{V}}_{n}$ is the volume
of ${\bf B}_{n}$. If it is large enough, the fundamental gravity scale
$M_{\star}$ may be as low as $\sim 1$ TeV \cite{add,add1}.

A key element of such higher-dimensional scenarios is the localization of
matter on   stable topological defects (branes \cite{polchinski},
domain walls \cite{local1,local2} or vortices \cite{add}; see also a recent
discussion \cite{DR}) embedded in
($4+n$)-dimensional bulk, with thickness $\lsim M_{\star}^{-1}$ and
surface tension $T\gsim M_{\star}^{4}$.

This paradigm offers a wealth of novel explanations
for the observed phenomenology --- patterns of supersymmetry and electroweak
symmetry breaking \cite{susygaugebreak}, three--genera\-tional structure \cite{gene},
ultra-light neutrinos \cite{neut}, proton stability \cite{proton}, etc.
It turns out that each particular basic aspect of phenomenology
explored so far is compatible with the brane--world ideas. Such a strategy ---
confronting established phenomenology with the brane--world ideas one by one ---
seems reasonable at the  present, exploratory stage.
The aspect which we address in this paper is a proliferation of distinct scales and
mechanisms in the current brane-world scenarios.
As simple brane-world ideas, that had been put forward several years ago,
were progressing and developing, they incorporated
contrived ``sub--mechanisms" and substructures, so that now there is an apparent
menace of producing a   ``personal" model for  each  particular phenomenon.
We pose a question whether it is possible to
find economic ways by combining several seemingly distinct mechanisms into one.
More concretely, we assume that at the primary stage
the matter sector of the theory has no mass parameters
whatsoever (i.e. purely conformal); the only mass parameters enters through gravity.
It then triggers a domain wall formation (determining its size and tension),
and electroweak symmetry breaking.

Conformal invariance --- the invariance of the physical laws under
rescalings of all lengths and durations by a common factor
 (see e.g. \cite{conf1,conf2}) --- is
broken in nature by particles' masses. If one starts from conformal matter, as we do,
it is concievable that mass parameters can penetrate
from gravity in two distinct ways.
First,  gravity loops  generate, generally speaking,
dimensionful constants in operators appearing in the Lagrangian for the
matter sector. This effect is not the one we are interested in.
We will ignore gravity loop corrections altogether.
This is a rather arbitrary assumption since we cannot indicate a dynamical pattern
 ensuring the required suppression of  the gravity loops.
Being aware that this is a weak point we will, nevertheless,
accept this assumption (quantum gravity is not a complete theory, anyway),
and will concentrate on another option. Treating gravity at the classical level,
we will ask the question:

\noindent
--- Is it possible that  the explicit breaking of conformal symmetry
in  the gravity sector (due to
the  bulk cosmological constant $\Lambda$ and other possible
sources) induces the
spontaneous breakdown of the conformal, gauge and other symmetries
in the matter sector
in an empirically viable way?

Certainly, such a scenario does not make any sense in
the context of the four-dimensional theories,
since in this case the fundamental gravity scale is given by $M_{\rm Pl} \sim 10^{19}$
GeV, while, say, the electroweak scale is $ \sim 10^{2}$ GeV.
However, in the context of the brane world scenarios, with
 low gravitational scales in the ballpark of  $M_{\star} \sim 10^{3}$ GeV,
the question above does not seem absurd. One may suspect that
the mass scales generated in the matter sector will
be of the same order of magnitude
 as the higher-dimensional fundamental  scale $M_{\star}$.

We will consider $n$ codimensions
and discuss  dynamics of a conformally-invariant gauge
theory on a factorizable manifold ${\bf M}_{4}\times {\bf B}_{n}$,
where the first factor represents our four-dimensional space-time,
while ${\bf B}_{n}$ represents extra dimensions and  is compact. It will be
arranged that the curvature scalar on  ${\bf B}_{n}$ will eventually
trigger the spontaneous breakdown of all symmetries in the matter sector. There are
three obvious requirements to be met. We must take care of: (i)
 the cosmological constant on ${\bf M}_{4}$; (ii)
the apparent gravitational scale on ${\bf M}_{4}$, i.e. $M_{\rm Pl}$;
(iii) the scale of the electroweak symmetry breaking, i.e.
the weak gauge boson masses. These requirements
 determine the size and curvature of
the internal manifold, as well as the bulk cosmological constant, in a correlated way.

Our construction  bears an illustrative nature.  As a reference point,
we will keep in
mind something like the Standard Model (SM).
However, we will focus mainly on  general aspects believing
that developing particular details would be premature at this stage.

The basic elements are as follows.
 To study the electroweak symmetry breaking we will
need to deal with a Higgs field.
To ensure that the matter sector is
described by an effective Lagrangian which is conformally invariant
we will need to introduce a universal dilaton field which replaces all dimensionful
SM parameters. Finally, we will need a ``defect builder" $\phi(x)$ (responsible for 
the formation of the topological
defect), on top of bulk gauge and fermion fields \cite{conf1,conf2}.

There is a certain conceptual similarity between the model we suggest and the
warped-compactification scenario of Randall and Sundrum \cite{rs1}.
In both cases the driving force is gravity.
This similarity does not go beyond the conceptual level, however.
In particular, the Randall-Sundrum model implies a single codimension,
which is not the case in our model. The hierarcy of scales is totally different too.

The organization of the paper is as follows. In Sec. 2
we consider a conformally-invariant gauge theory in
the factorizable geometry ${\bf M}_{4}\times {\bf B}_{n}$.
After
specifiying the properties of ${\bf B}_{n}$ required in order to get
appropriate
phenomenology, we discuss the emergence of $M_{\rm Pl}$ form $M_{\star}$,
the formation of the  topologically stable defect and  matter localization on the defect.
In Sec. 3 we discuss stability of the factorizable
geometry. In particular, we check the consistency of the static
background by tuning the long-distance cosmological constant to zero
while keeping the tension of order of  $M_{\star}^4$. In Sec. 4 we summarize our
conclusions and comment on similarities/distinctions with other popular brane-world
scenarios.

\section{ Conformal gauge theories on compact hyperbolic manifolds}

The framework of our discussions is a ($4+n$) dimensional static factorizable
geometry ${\bf M}_{4}\times {\bf B}_{n}$ where ${\bf B}_{n}$ is a
compact manifold and ${\bf M}_{4}$ is the ordinary (empirically flat)
space-time. In the static limit, generically,
 the compact manifold can have positive
 (e.g. an $n$-sphere ${\bf S}_{n}$), vanishing (e.g. an $n$-torus ${\bf T}_{n}$),
or negative  (e.g. an $n$-dimensional compact hyperbolic space) curvature
scalar,  depending on its geometry and topology \cite{kaloper,cancel}.
Compact negative scalar curvature manifolds can be obtained from a noncompact one
by applying a known procedure, see below. It is also necessary to verify the
stability of the chosen background geometry. This issue will be discussed in Sec. 3.

As was already mentioned, we will require the nongravity part of the theory
of the SM type  to be
conformal in the $(4+n)$--dimensional space. The conformal invariance is achieved
through the dilaton coupling. Then, the matter part of the stress tensor
is strictly traceless \footnote{Here we
neglect possible conformal anomalies which are not expected to play
a role;  moreover, for odd $n$ (the case $n=3$ may be preferrable for our purposes)
they are absent.}.
The traceless nature
of the stress tensor is particularly important for us since  it implies that the curvature
scalar, ${\cal{R}}$, is entirely determined from the
classical gravity equations by the bulk cosmological term and
the background geometry,  independently of the matter sector dynamics.

In general, the conformal invariance
puts severe restrictions on possible couplings of the matter fields \cite{conf1,conf2}.
One key aspect is that the matter sector can contain no mass parameters ---
they  can be
generated only via  gravitational interactions. At tree level, the
matter  sector couples to  gravity via the minimal coupling to
the metric field, and via the conformal coupling ${\cal{R}} \sigma^{2}$
of a  scalar field $\sigma$. It is this latter coupling that  is particularly important as it
induces mass terms for scalars in the constant-curvature background. Neglecting the
curvature of ${\bf M}_{4}$ (as it will eventually be tuned to zero) one concludes that
a conformal scalar is either massive (${\cal{R}}>0$),  masless (${\cal{R}}=0$),
or tachyonic (${\cal{R}}<0$) depending on the structure of ${\bf B}_{n}$.

Our  goal is generation of the observed particle spectrum from a conformal 
higher-dimensional gauge theory.  It is clear then  that the induced breaking of the
conformal invariance in the matter sector must generate an instability in the vacuum state.
Obviously, for this to happen, it is necessary to have a negative
curvature scalar (this may not be sufficient, as will be discussed below).

In general, a smooth compact manifold ${\bf B}_{n}$ of constant negative curvature is
obtained from the covering space ${\bf H}_n$  of $n$-dimensional hyperbolic spaces by
modding out by a freely and discontinuously acting (with  no fixed points) subgroup
$\Gamma$ of its isometry group. Therefore, hereon we take the internal
manifold to be ${\bf B}_{n}={\bf H}_n/\Gamma$,
 with the constant negative curvature
${\cal{R}}_{0}$. This is a highly curved negative-curvature manifold with a
global anistropy and  rigidity (no massless shape moduli) \cite{kaloper,others}. The
volume of such manifolds  grows exponentially with their linear size, and it is the
largest linear extension $L$ that dominates
\begin{eqnarray}
{\cal{V}}_n = |{\cal{R}}_{0}|^{-n/2} e^{(n-1) \sqrt{|{\cal{R}}_{0}|} L}
\label{two}
\end{eqnarray}
in $n\geq 2$ codimensions. Here it is assumed that  $|{\cal{R}}_{0}|^{1/2}  L\gg 1$ and
we  neglected irrelevant
angular factors in Eq. (\ref{two}).

The graviton zero mode on such
manifolds is a constant
\cite{kaloper}, and, therefore, the  hierarchy problem is solved by virtue
of  their large volume. Combining Eqs. (\ref{one}) and (\ref{two})
we get
\begin{eqnarray}
\label{hier}
\frac{M_{\rm Pl}^{2}}{M_{\star}^{2}}=e^{(n-1) \sqrt{|{\cal{R}}_{0}|} L}
\left(\frac{M_{\star}^{2}}{|{\cal{R}}_{0}|}\right)^{n/2}\,.
\end{eqnarray}
A huge hierarchy between $M_{\star}$ and $M_{\rm Pl}$ is generated by the
topological invariant $\exp{[(n-1) \sqrt{|{\cal{R}}_{0}|} L]}$ in ${\cal{V}}_{n}$.
Since the dependence on $L$ is exponential,
unlike in the original proposal \cite{add,add1},
one can settle for a microscopic size of the compact manifold in the
extra dimensions.
Clearly,
the fundamental scale of gravity $M_{\star}$ does not need to exactly coincide with the
scale of $|{\cal{R}}_{0}|$ (and also with the bulk cosmological constant $\Lambda$, see
below). In fact, one can choose $|{\cal{R}}_{0}|
\simlt M_{\star}^{2}\sim ({\rm  TeV})^{2}$ by adjusting $\sqrt{|{\cal{R}}_{0}|} L$ appropriately. For instance,
if $$n=3\,, \quad M_{\star}\sim  1\ {\rm TeV}\,,\quad {\rm and}
\quad |{\cal{R}}_{0}|\sim (0.5\
{\rm TeV})^{2}\,,$$
 the maximal linear extension  of the manifold turns out to be
$$L\approx 34\ |{\cal{R}}_{0}|^{-1/2}\approx 1.4 \times 10^{-15}\ {\rm cm}\,.$$
Unlike the Arkani-Hamed--Dimopoulos--Dvali 
(ADD) scenario \cite{add} where the size of the extra dimensions is
macroscopic (and is at the border of what is  allowed by the current
gravity experiments \cite{exp}, $\sim 0.1\ {\rm mm}$)
in the case at hand $L$ is microscopic.

\subsection{Relation between the scalar curvature and the bulk cosmological term}

In a higher-dimensional theory whose nongravity part is strictly conformal,
the Einstein equations imply that the curvature scalar is determined solely
by the vacuum energy densities (the bulk cosmological constant
plus other possible sources). For a factorizable geometry ${\bf M}_{4}\times {\bf B}_{n}$,
assuming that ${\bf M}_{4}$ is already flattened thanks to appropriate source terms,
the compact space possesses the curvature scalar ${\cal{R}}= 2 n \Lambda/(n-2)$
where $\Lambda$ is the bulk cosmological term. For grasping the importance of the
static character of the internal manifold, one notices that the curvature
scalar has the form $${\cal{R}}={\cal{R}}(d^2r/dt^2, (dr/dt)^{2}, r^2)\,,$$
$r$ being the curvature radius of the internal manifold \cite{cancel}.
Clearly, for ensuring a static compact space, the intrinsic curvature
contribution (the only piece independent of the time derivatives) to
${\cal{R}}$ must be balanced by the bulk cosmological term in the field
equtions: ${\cal{R}}\propto \Lambda$.

In analyzing the matter sector, we take the factorizable static background geometry
as the basic {\em ansatz}. The consistency of this assumption as well as the relation
between the curvature scalar and the bulk cosmological term are best understood
after reducing the bulk field theory to ${\bf M}_{4}$, and requiring the
stability and vanishing of the long-distance (four-dimensional)
cosmological term. Such
details are deferred till Sec. 3.

\subsection{ Destabilization of scalar fields}
It is convenient to discuss first the destabilization of
the scalar potential of a typical scalar field. For instance, a dilaton $\sigma$
may be described by the Lagrangian
\begin{eqnarray}
\label{dilat}
{\cal{L}}[{\cal{R}},\sigma] &=& (1/2)\left[G^{A B} \partial_{A} \sigma \partial_{B} \sigma
-\zeta_c {\cal{R}}_0 \sigma^{2} - \lambda_{\sigma} \sigma^{2 \gamma}\right]
\end{eqnarray}
where $\sigma$ is a real field,
$$
\zeta_c=\frac{n+2}{4 (n+3)} \quad {\rm and}\quad  \gamma= \frac{n+4}{n+2}\,,
$$
as required by the conformal invariance \cite{conf1,conf2}. The potential of $\sigma$,
\begin{equation}
V(\sigma)=\zeta_c {\cal{R}}_0 \sigma^{2} + \lambda_{\sigma}
\sigma^{2 \gamma}\,,
\end{equation}
 has two\footnote{A minimum at negative $\sigma$ is irrelevant
for our purposes.} critical points $\sigma_{\rm max}=0$  with
$V(\sigma_{\rm max})=0$, and
\begin{eqnarray}
\sigma_{\rm min}&=&\left(-\frac{\zeta_c {\cal{R}}_0}{\gamma
\lambda_{\sigma}}\right)^{1/2(\gamma-1)}\,,\nonumber\\[0.2cm]
 V(\sigma_{\rm min})&=&\left(\frac{1-\gamma}{2 \gamma}\right) \left(\gamma
\lambda_{\sigma}\right)^{-1/(\gamma-1)}
|\zeta_c{\cal{R}}_0|^{\gamma/(\gamma-1)}\,,
\end{eqnarray}
which correspond to local maximum and minimum, respectively. The small
 perturbations $\overline{\sigma}$ around
these critical points  have masses $m_{\sigma}^{2}({\rm max})=\zeta_c {\cal{R}}$ and
$m_{\sigma}^{2}({\rm min})=2 (1-\gamma)
\zeta_c {\cal{R}}_0$.
Remember that the quantities ${\cal{R}}_0$ and $1-\gamma$ are negative.

The  $\sigma$ quanta evolve in time as $\overline{\sigma}\sim e^{i m_{\sigma} t}$
which  implies that small perturbations  around $\sigma_{max}=0$ are
unstable.
Since ${\bf B}_{n}$ is a compact manifold, one always has  a zero mode solution
$\overline{\sigma}(x,y)={\rm const}\  \overline{\sigma}_{0}(x)$ where $y$ stands for extra
coordinates, and $\overline{\sigma}_{0}(x)$
 obeys the equation
$$\Box_{4} \overline{\sigma}_{0}(x)+ \zeta\ {\cal{R}}_0\ \overline{\sigma}_{0}(x) =0$$
 whose solution is
always destabilized. Note that a perturbative stability analysis of Ref.
\cite{ads} referring to noncompact anti-de Sitter spaces
is inapplicable in the case at hand due to the
compactness of
${\bf B}_{n}$. The vacuum expectation value of $\sigma$ is necessarily
nonvanishing for a negatively-curved internal manifold. This is a spontaneous breaking
 effect which will
communicate the explicit conformal symmetry
breaking of the gravity sector to
the matter sector.

\subsection{``Defect builder" and the Higgs fields}

Having discussed the destabilization of a typical scalar field via 
constant negative curvature scalar, we now turn
to the issue of localization of matter at distances $\lsim M_{\star}^{-1}$
on (empirically flat) submanifold ${\bf M}_{4}$. (The bulk theory also  
possesses gauge and other
symmetries, to be spontaneously  broken; this will be discussed later).
 There are various field-theoretic
and  stringy mechanisms for
localizing matter on ${\bf M}_{4}$. For our illustrative
purposes we will  utilize a field-theoretic framework put forward and
 developed in \cite{local1,local2} in which the ordinary four-dimensional
space-time is a topologically  stable defect. This is by no means a unique option.
One could consider other known mechanisms leading to matter localization.

In general, the
formation of the stable defect requires a spontaneously broken global symmetry.
Moreover, the
type of the defect depends on the number of extra dimensions: a domain wall in one codimension,
a vortex line in two codimensions, and so on.

 Unlike the spherical or toroidal structures \cite{add}, the
manifold ${\bf B}_{n}$ under consideration  is globally anistropic \cite{others}; for solving the
hierarchy problem only the  largest linear size $L$ is relevant \cite{kaloper}.
Therefore, as an
approximate  but physical picture,  one can imagine ${\bf B}_{n}$ extending along a particular direction,
say $y$, like a stick
\footnote{ The shape and size of the manifold depends on what subgroup of the isometry group of ${\bf
H}_{n}$ is acting. For a detailed numerical study of ${\bf B}_{4}$ see \cite{others}.}
of
length $L$ and thickness $\delta\ll L$,
$$
\delta \sim |{\cal R}_{0}|^{-1/2}\,,\quad L/\delta \gg 1\,.
$$
In other words, out of all $n$ dimensions, one is significantly
larger than  the remaining $n-1$. The latter, though needed to keep the
curvature scalar negative and constant, are much smaller. Clearly,
within such a picture, the dependence  of matter fields on these
$n-1$ dimensions can be neglected (as well as the corresponding components of, say, vector fields).
The problem
becomes effectively  one-dimensional. In such  quasi one-dimensional
setting, the defect builder field
$\phi$ can form a domain wall and dynamically localize the matter \footnote{ The topological
stability of  the domain wall requires an infinite extension for $y$, and therefore, the
picture discussed here is  approximate;  the wall
will be approximately stable. Its decay rate will be suppressed
exponentially as $\exp (-{\rm const} L |{\cal{R}}_0|^{-1/2})$. }.
Consequently, the scalar sector, composed of the defect builder $\phi$, dilaton
$\sigma$ and the Higgs field $H$,  may be described by the Lagrangian
\begin{eqnarray}
\label{scal}
{\cal{L}}[{\cal{R}}, \phi, \sigma, H] &&=(1/2)\left[
G^{A B} \partial_{A} \phi \partial_{B}\phi-\zeta_c {\cal{R}} \phi^{2} - \lambda_{\phi} \phi^{4} \sigma^{2
(\gamma-2)}\right]+\nonumber\\[0.2cm]
&& G^{A B} ({\cal{D}}_{A} H)^{\dagger} {\cal{D}}_B H
-\left(\zeta_c {\cal{R}} + \lambda_{0} \phi^{2}\sigma^{2 (\gamma-2)}\right) H^{\dagger} H
- \lambda_{h} \left(H^{\dagger}H\right)^{\gamma}
\end{eqnarray}
to which (\ref{dilat}) is to be added.  Here we take $\lambda_{\phi}$ to be sufficiently 
small compared to $\lambda_{\sigma}$ so that the change in $\sigma_{min}$ is small.  For
simplicity one can assume $\phi$ to be real. The interactions of the scalars are such that
there  is a manifest ${\cal{Z}}_{2}$ invariance under which $\phi\rightarrow -\phi$, 
$\sigma\rightarrow \sigma$, and $H\rightarrow H$. 

In Eq. (\ref{scal}) we  dropped several terms allowed by symmetries as such
terms are not essential
for  the mechanism discussed here. For instance, in the Higgs interaction one can add
terms $\phi^{4} \sigma^{2 (\gamma-3)} H^{\dagger} H$ and $\sigma^{2 (\gamma-1)} H^{\dagger} H$, whose main effect
would be to split the
scalar masses. In any case, as we are not aiming at reproducing the exact electroweak spectrum, such
details are not essential.
 
In the $\sigma=\sigma_{\rm min}$ background, the potential of $\phi$ is destabilized, leading to a
spontaneous breakdown of the ${\cal{Z}}_2$ symmetry with two possible VEVs,
\begin{eqnarray}
\phi_0=\pm \varphi_{0}\,,\qquad
\varphi_{0} = \left(\frac{ |\zeta_c {\cal{R}}_0|}{2 \lambda_{\phi} \sigma_{min}^{2(\gamma-2)}}
\right)^{1/2}\,.
\end{eqnarray}
This allows one to build a wall
with the profile
\begin{eqnarray}
\phi(y)=\varphi_{0} \tanh \left( m_{\phi} y\right)
\end{eqnarray}
interpolating between $- \varphi_{0}$ and $+ \varphi_{0}$  as $y$ changes from
$-L/2$ to $L/2$.
Here $m_{\phi}^{2}=|\zeta_c {\cal{R}}_0|$ is the mass of the $\phi$ quantum.
Needless to say that $m_\phi$ is assumed to be large, $m_{\phi} L \gg 1$.
Then the wall thickness is much less than $L$. The inverse thickness of the
wall  as well as its tension ($T\sim |\zeta_c {\cal{R}}_0|^{2}$) are in the
ballpark of $M_{\star}$ to the appropriate power (see Fig. 1).

We now turn to the discussion of the Higgs field in the domain wall background.
With $\lambda_{0}>0$ the term $\phi^{2} \sigma^{2(\gamma-2)} H^{\dagger} H$
induces a positive potential for $H$ in the bulk, while 
in the core of the defect the potential vanishes. Then the Higgs field is stable 
outside the wall,
while
 a tachyonic mass   develops inside the wall, and, hence, 
a nonvanishing VEV of the Higgs field develops inside the wall \footnote{The impact of
the gauge interactions of the Higgs field on stabilization/destabilization 
of the Higgs potential will be discussed
in Sect. 2.4. It does not change the overall picture.}.   More concretely,  away from the
wall, in the bulk, the wall builder attains one of its two vacuum values, and the effective
(mass)$^{2}$ of the Higgs field is
\begin{eqnarray}
\widetilde{m}_{H}^{2}= \{- 1 + \lambda_{0}/(2 \lambda_{\phi})\}
|\zeta_c {\cal{R}}_0|\,.
\label{f20}
\end{eqnarray}
The mass term (\ref{f20}) is
positive provided  $\lambda_{0}>2
\lambda_{\phi}$, and, consequently, the gauge symmetry remains unbroken in the bulk.
In the core of the wall
\begin{eqnarray}
\widetilde{m}_{H}^{2}\approx  -  
|\zeta_c {\cal{R}}_0|\,,
\label{f21}
\end{eqnarray}
and the gauge symmetry is broken provided 
$\lambda_{0}>2
\lambda_{\phi}$. 
The constraint $\lambda_{0}>2
\lambda_{\phi}$  is a mild tuning, and such a choice does not produce any harm on the
mechanism of the wall formation.

In the core of the domain wall, however, $\phi(y)\sim 0$, and thus, the Higgs field
necessarily develops a
nonvanishing VEV
\begin{eqnarray}
\left|H\right|_{0}=\left(\frac{\zeta_c |{\cal{R}}_0|}
{\gamma\lambda_h}\right)^{1/2(\gamma-1)} \,,
\end{eqnarray}
which
leads, in turn,  to a spontaneous breakdown of the gauge symmetry. In this
minimum, the mass of the
Higgs quantum is given by $m_{h}^{2}= 4 (\gamma -1) |\zeta_c {\cal{R}}_0|$.
As mentioned above, had
we included terms like $\phi^{4} \sigma^{2 (\gamma-3)}$ and
$\sigma^{2 (\gamma-1)}$ the
appearent degeneracy between $\sigma$ and $H$ quanta would be lifted,
and the condition on
$\lambda_0$ to avoid breaking of the gauge symmetry in the bulk would be
also modified accordingly.

\subsection{Gauge fields}

To outline the gauge field dynamics let us consider an SU(2) gauge theory in the
bulk with
$${\cal{D}}_A H = \left(\partial_A +i (g_2/2)
\widetilde{\sigma}\vec{\sigma}\cdot \vec{W}_{A}\right)H$$
where $\vec{W}_{A}$ are the three  gluon fields with $4+n$ components, 
and $\widetilde{\sigma}$ is a dilaton with mass dimension $-n/2$,
\begin{eqnarray}
\widetilde{\sigma} = -\left( 1+\frac{2}{n}\right)\,\sigma^{-n/(n+2)}\,.
\label{f22}
\end{eqnarray}
For the stick-like manifold configuration under consideration, $\vec{W}_{A}$ is 
effectively a five-dimensional gauge field. Then, in the core of the wall, 
the SU(2) symmetry is completely broken giving three massive vector bosons,
\begin{eqnarray}
M_{W}^{2}  = \left(\frac{g_2}{\gamma -2}\right)^{2}
\left(\frac{\lambda_{\sigma}}{\lambda_{h}}\right)^{1/(\gamma -1)}
\frac{\left|\zeta_c {\cal{R}}_0\right|}
{\gamma \lambda_{\sigma}}
\end{eqnarray}
whose degeneracy can be lifted  by additional group factors (e.g. the hypercharge group
U(1)$_{Y}$). Outside the core of the wall the Higgs field does not condense, and the gauge
theory remains in the non-Abelian   phase. One can try to exploit this fact
in order to use
 the mechanism
for the gauge field localization  on the wall suggested in Ref. \cite{local2}.

The   mechanism   \cite{local2}
requires that in  the bulk we have an unbroken  non-Abelian gauge theory,
which develops string-like flux tubes with 
a nonvanishing  string tension. Then,
just like in the Meissner effect (where a perfect superconductor repels 
the magnetic field), the ``superconducting"  bulk (where the wall--builder 
condenses keeping the gauge invariance  exact) will repel the flux tubes 
of the electric field confining them to the core of the topological defect. 

The question whether or not  the flux tubes are formed
and the non-Abelian theories confine in $4+n$
dimensions is not completely clear. We do know that they confine in two, three and four
dimensions. Moreover, at sufficiently strong coupling confinement persists
in higher dimensions, as follows from lattices and from ADS/CFT-based arguments
\cite{PS} (see also \cite{add} for similar field-theoretic arguments). In the D-brane/string theory it is explicit.
Thus, it is likely that in a certain range of the coupling constants $\lambda$
the effective
coupling constant
$\widetilde{\sigma} g$ is such that gauge-nonsinglet objects are confined
in the bulk at $n>0$.  Then the mechanism of Ref. \cite{local2}
applies. A particular model of how this mechanism localizes gluons was considered in 
Ref. \cite{DR} in detail. The problem is whether or not one can
get  a sufficiently small gauge coupling (compatible with phenomenology)
inside the wall, where the gauge symmetry is spontaneously broken, and we 
are in the Higgs
phase. To this end one can try to play with the variations of $\widetilde{\sigma}$
inside and outside the wall. This task would require a numerical analysis, which we 
postpone for the future.

In Sect. 2.3 we have discussed the bulk stabilization (core destabilization) of the
Higgs potential with the gauge interactions switched off (see Eqs. (\ref{f20}) and
(\ref{f21})). Here we will argue that the phenomenon remains valid with the gauge
interactions switched on. The gauge interactions induce a contribution to the Higgs mass
term of the form
\begin{eqnarray}
\langle H^\dagger (-D^2_{\rm Euclid})H\rangle  \to C M_*^2 H^\dagger H\,,
\end{eqnarray}
where $D^2_{\rm Euclid}$ is the Euclidean covariant Lapacian,
and $C$ is a dimensioneless constant which depends on various other dimensionless
constants in the theory, such as $\lambda_\sigma$. 
It is important that because of positivity of
$-(D^2_{\rm Euclid})$, the constant $C$ is positive. It adds to the stabilizing term in the
Higgs potential. If $C$  is adjusted to be not too large,
the previous conclusion of  the bulk stabilization and core destabilization 
remains intact.

The above dynamical trapping is not special to gauge fields. In fact, the bulk 
fermions will be localized on the wall too due to  the confining gauge 
dynamics outside. For the purpose of localization, they may or may not be directly coupled  to
the wall  builder as in \cite{local1,DS}).
 As a simple example, consider an  SU(2)
doublet $\psi$ and an SU(2)
singlet $\psi^{\prime}$ with the Yukawa interaction
\begin{equation}
\label{ferm}
{\cal{L}}_{Y}= y_{\psi} \widetilde{\sigma} \overline{\psi} H
\psi^{\prime} + \mbox{h. c.}
\label{wed}
\end{equation}
Then in the core of the wall, where  the SU(2) symmetry is spontaneously
broken, there arises a massive fermion with mass
$m_{\psi}=y_{\psi} M_{W}/\sqrt{g_2}$.
To see how realistic this mass spectrum is,  we take
$n=3$, $\lambda_{\sigma}\sim \lambda_{h}\sim 1/\gamma
\ll 1$ and $|{\cal{R}}_0|\sim (350\ {\rm GeV})^{2}$.
This then gives $M_{W}\sim 100\ {\rm GeV}$,
$m_{\psi}\sim y_{\psi} M_{W}$, and $m_{h}=200 \ {\rm GeV}$
where one particularly notices that
the Higgs boson is always heavier than the gauge bosons.
   
An important point to be addressed here is the chirality of the localized
fermions. In general, fermions localized via the confining bulk gauge
dynamics are not chiral. Moreover, in odd dimensions ($n$ is odd) there is no chirality.
In general \cite{local1,local2},  a way out
is provided by the defect--builder itself: if  the bulk fermions are
  directly coupled to $\phi$, the zero modes are chiral, and
their chirality is correlated with the topological charge of the
wall. Explicitly, the bulk fermions $f$
and
$f_c$ can be coupled as $y \widetilde{\sigma} \phi \overline{f} f + y_c \widetilde{\sigma} \phi
\overline{f_{c}}  f_{c}$ which deposit the chiral zero modes $f + \overline{f}^{\dagger}$
and $f_c^{\dagger} + \overline{f_c}$ whose couplings to the Higgs field
as in (\ref{ferm}) produces   acceptable fermion masses in the 
core of the defect \cite{add}.

Concerning the gauge symmetry in the bulk,  one should take care of gauge 
singlets, as $\psi'$ in Eq. (\ref{wed}). To this end one may embed the SM 
gauge group in a larger non-Abelian one, $e.g.$ the Pati-Salam group SU(4)$\times$ SU(2)$_L\times$ SU(2)$_R$. 
Then using appropriate (in number and representation) Higgs fields one can 
obtain the SM spectrum below $M_{\star}$. In fact a realistic model with 
Pati--Salam group in $n=2$ codimensions have already been discussed in 
\cite{add} where the defect builder forms a vortex line the throat of which 
consistently localizes the matter on ${\bf M}_{4}$. One notices that once the 
scalar sector is destabilized, the hierarchy needed among VEVs 
of various Higgs fields can be generated via their interactions by mild tuning of
the parameters (e.g. couplings of the form $\phi^{4} \sigma^{2 (\gamma-3)} H^{\dagger} H$).

\section{Stabilization of the factorizable geometry and flatness of ${\bf M}_{4}$}

The discussion in Sec. 2  was based on the factorizable background geometry
${\bf M}_{4}\times {\bf B}_{n}$ where ${\bf B}_{n}$ is a static compact
negative-curvature manifold.  Here our primary concern
is the stabilization of the extra dimensions. Let us recall, for instance,
that in spherical or toroidal geometries \cite{add,cancel,cancelp} one has
to stabilize the large extra dimensions against
expansion, which requires  the bulk cosmological constant be balanced with the curvature
of the manifold, and against  contraction, which requires, generally speaking, either
brane-lattice  crystallization or a topological invariant, $e.g.$ Ramond-Ramond gauge field
on ${\bf S}^{2}$ topology.

For the manifold structure under consideration, the main problem is to prevent the
expansion of the internal manifold as its size is already required to be around the
fundamental scale of gravity. As in positive-  or zero-curvature spaces
\cite{cancel,cancelp} the stabilization against the expansion requires fine-tuning  the bulk
cosmological constant against the curvature term. Indeed, the analysis of \cite{kaloper}
shows that a factorizable geometry of the form ${\bf M}_{4}\times {\bf B}_{n}$ almost
automatically arises once the bulk cosmological constant is appropriately tuned.

The compact hyperbolic manifolds  possess the important property of rigidity ---
their volume in units of $|{\cal{R}}|$ cannot be changed while maintaining the
homogeneity of the space. Therefore, stabilization of the static factorizable background
reduces to the stabilization of the curvature length $|{\cal{R}}|^{-1/2}$ of ${\bf B}_{n}$.
The relevant part of the reduced bulk action in the far infrared is nothing but
the long-distance (four-dimensional) cosmological constant,
\begin{eqnarray}
\label{4cc}
\Lambda_{4}({\cal{R}}) = {\cal{V}}_{n}\left(-M_{\star}^{n+2}({\cal{R}} - 2 \Lambda) +
V(\sigma_{min})\right) + {\cal{K}}({\cal{R}}) + T
\end{eqnarray}
where $T>0$ is the wall tension (including the contribution of the Higgs potential). Here
${\cal{K}}({\cal{R}})$ collectively denotes the contribution of the kinetic
terms of the bulk scalar and vector fields \cite{cancel,cancelp}, and can be expanded as
$\sum_{a>0} C_{a} |{\cal{R}}|^{a/2} M_{\star}^{4-a}$ \cite{kaloper}. The stability
of the factorizable configuration requires that
$$\Lambda_{4}^{\prime} \left({\cal{R}}={\cal{R}}_{0}\right) = 0 \quad{\rm and}
\quad \Lambda_{4}^{\prime\prime}
\left({\cal{R}}={\cal{R}}_{0}\right) > 0\,,$$
together with the empirical requirement of $\Lambda_{4}\left({\cal{R}}={\cal{R}}_{0}\right) = 0$.
A straightforward calculation, which is particularly simple for large $n$, suggests that
$\Lambda<0$ and ${\cal{R}}\equiv {\cal{R}}_{0}\sim 2 \Lambda$. Moreover, $\Lambda_{4}^{\prime\prime}
\left({\cal{R}}_{0}\right)$ determines the masses of small fluctuations around
${\cal{R}}={\cal{R}}_{0}$ to be ${\cal{O}}(M_{\star})$ which is large enough to
evade cosmological problems \cite{kaloper} with a light radion occuring in spherical
and toroidal geometries \cite{cancel}. One notices that the bulk cosmological
constant prevents the internal manifold from  expanding  indefinitely.

An important issue in the far infrared is the vanishing of the long-distance cosmological
constant. Empirically, it is known that such a cancellation can be achieved by tuning the
coefficients $C_a$ against the first and third terms in (\ref{4cc}) which involves
an extreme fine-tuning of the parameters --- the cosmological constant problem.
Clearly, one should excercise care in fine-tuning $\Lambda_{4}({\cal{R}})$ to zero
in order to make ${\bf M}_{4}$ flat, as it can result in a
solution with $T\sim {\cal{V}}_{n} M_{\star}^{n+4} \gg M_{\star}^{4}=
 M_{\rm Pl}^{2} M_{\star}^{2}$
which is completely unnatural given the characteristic scale of the SM. Moreover, such
a huge wall tension will destroy the initial ansatz on the background geometry.
On the contrary, if $T\sim M_{\star}^4$, the back reaction on the wall
on the solution under consideration is negligible,
and our step-by-step strategy is justified.

One can estimate the back reaction of the wall by examining the
Einstein equations. In the presence of the wall a new term in the right-hand side
appears, proportional to $T\delta^n(\vec y)$. In a rough approximation we will
replace $T\delta^n(\vec y)$ by $T / {\cal V}_n$, smearing the delta function homogeneously
over the extra space. This will presumably lead to an overestimate of the back reaction.
Neglecting  irrelevant numerical factors, we get
\begin{eqnarray}
\Delta {\cal{R}} \sim  \frac{T}{M_{\rm Pl}^{2}}\,.
\end{eqnarray}
Let us remind the reader that this is a purely classical estimate, with
all quantum corrections discarded. Moreover,  $\Delta {\cal{R}}$
need not be constant, it depends on the profile of the wall
solution in $\vec y$. It is clear that
when $T\sim {\cal{V}}_{n} M_{\star}^{n+4}=M_{\rm Pl}^{2} M_{\star}^{2}$
the change in ${\cal{R}}_{0}$ is ${\cal{O}}(1)$, that is, the original
ansatz on the curvature of the space is completely
destroyed, and our construction collapses. Therefore,
to prevent the ambient geometry from being significantly modified by
the back reaction of the wall, one must require $T\simlt M_{\star}^{4}$.
This constraint represents a nontrivial aspect of the
long-distance cosmological constant problem \cite{cancel,cancelp,kaloper}.

\section{Summary and Discussion}

\begin{figure}[t]
\begin{center}
\vspace{9pt}
\epsfig{file=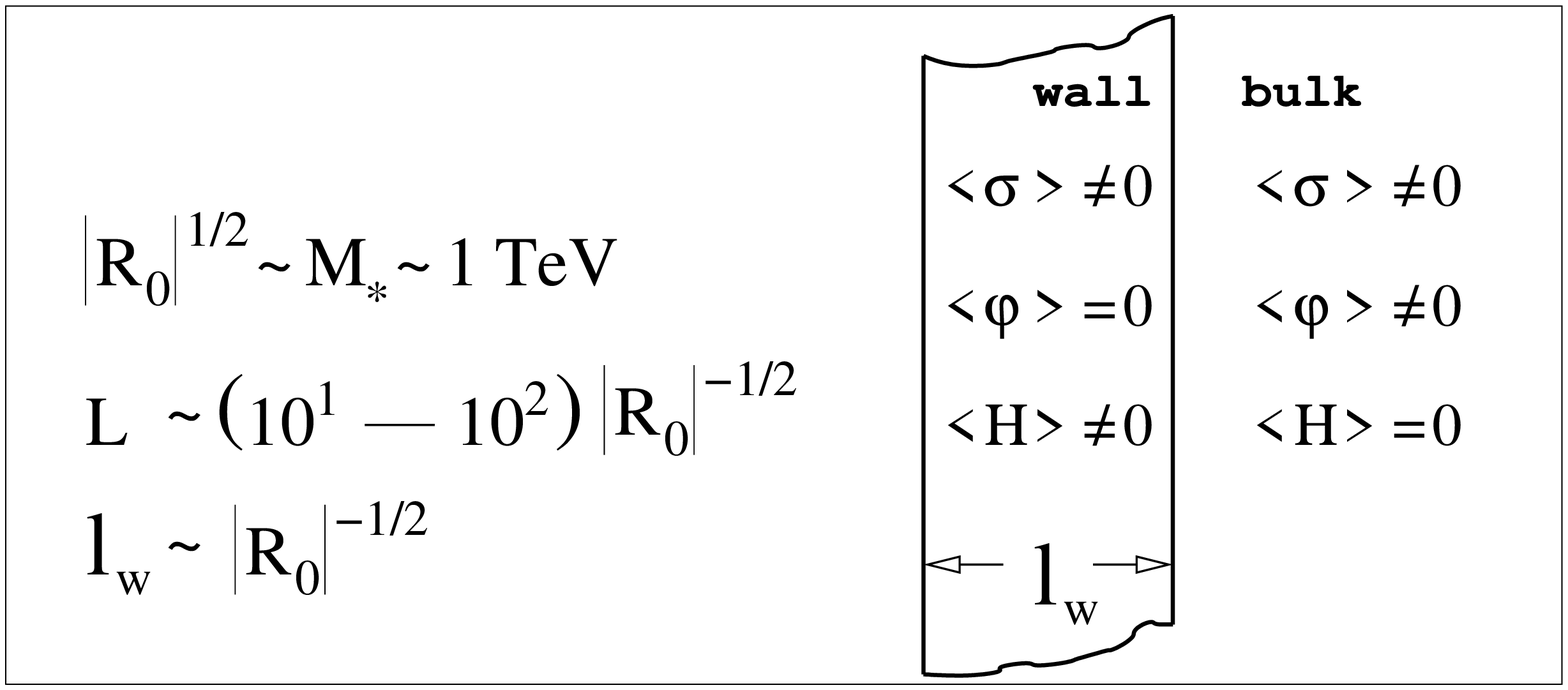,height=2.5in}
\caption{\label{scales}
{\it List of scales in the problem together
with the VEVs of fundamental scalars in 
the bulk and the wall.}}
\end{center}  
\end{figure}

Our basic starting point is the assumption that  large compact
extra dimensions exist and, as a result, the fundamental
gravity scale $M_{\star}\sim {\rm TeV}$. Unlike the original suggestion
\cite{add} where most attention is paid to flat extra dimension,
a (constant) negative curvature of the extra space is absolutely essential for the
mechanism that we discuss. It is maintained, in turn, by a bulk negative cosmological term.

In this paper we outline a single-origin step-by-step mechanism
which might be relevant for the low-energy phenomenology. 
Initially, the mass parameters are separated --- the nongravity sector is
assumed to be conformal. A (negative) cosmological term in the bulk
ensures the existence of the negative curvature internal compact manifold.
Its scalar curvature is related to the bulk cosmological term.
It triggers then the formation of a topological defect, which, in turn,
captures the SM matter fields. Simultaneously, it destabilizes the
Higgs potential and triggers the spontaneous breaking of the gauge symmetry on the the defect
(but not in the bulk).  The scales of the problem and VEVs of the
fundamental scalars are summarized in Fig 1.  Clearly, all 
mass and length scales are fixed by the curvature scalar ${\cal{R}}_0$.

The stability of the background
geometry is consistent and self-sustaining
provided that the bulk cosmological constant
$\sim - M_{\star}^{2}$. The scale of gravity $M_{\star}$, the curvature scalar
${\cal{R}}_{0}$, bulk cosmological term $\Lambda$, and the largest linear extension of the
compact manifold $L$ are the  mass parameters of the model. These mass scales are
interrelated via the requirements of (i) generating the correct electroweak spectrum
($|{\cal{R}}_{0}|^{1/2}\sim$ Higgs mass); (ii) explaining enormity of $M_{\rm Pl}$ with
respect to electroweak scale, and (iii) cancellation of the four-dimensional cosmological
constant. A modest hierarchy is required between
$L$ and $|{\cal R}_0|^{-1/2}$,
$$L |{\cal{R}}_{0}|^{1/2}\gsim 30\,,$$
which may be purely numerical.

Recent studies of the cosmology of compact hyperbolic manifolds \cite{cosmo} suggest 
that the topology of extra space implies a very large entropy, and statistical
averaging during collapse can account for the large scale flatness. However, in
both publications \cite{cosmo} and \cite{kaloper} nongeometric degrees of freedom (in the 
bulk and on the defect) have not been included in the discussion. Analysis of
several problems, $e.g.$ CP violation, baryogenesis, inflation, structure formation $etc.$,
would be in order along the lines we have outlined in the present paper. 

A toy SU(2)  gauge model illustrates that the existence of a
negatively-curved compact internal manifold alone is sufficient for  breaking the
symmetries of the matter sector so as to generate a flat topological defect trapping the
massive as well as massless particle spectrum on it.
Supersymmetry may or may not be needed. One can consider
models with supersymmetry broken by the bulk cosmological term along these lines.

To conclude, it would be in order to compare the pattern suggested here
with other popular  brane-world scenarios \cite{add,rs1}.
The compactness of the extra space is the basic feature of the ADD scenario.
However, the characteristic size of
the extra dimensions for compact hyperbolic structures  is far below
 those obtained in spherical or toroidal 
geometries \cite{add} (and orders of magnitude  below the existing experimental bounds
\cite{exp}). Small extra dimensions  occur   in the warped compactification   scenario
of Randall and Sundrum
\cite{rs1}. The two approaches have similarities and distinctions.
The basic conceptual similarity is the fact
that in the both schemes gravity in the bulk plays a crucial role.
The distinctions are technical but conspicuous.
To name a few one
notes that: ($i$) the warped compactification   scenarios work  for a single extra dimension
whereas the scenario  discussed here requires at least two extra dimensions; ($ii$) In the
warped compactification   scenarios
five-- and four--dimensional Planck scales essentially coincide while
they are drastically different in 
the case of compact hyperbolic structures; ($iii$) In both 
scenarios the bulk cosmological constant is required to be negative,
but its scale is  different. As is seen from Fig. 1,
the basic and practically the only scale in our model is $M_*\sim |{\cal R}|^{-1/2}$.
In the Randal-Sundrum approach one starts from the $M_{\rm Pl}$ brane;
($iv$) The gravity-driven trigerring of the Higgs mechanism on the TeV brane 
is not a part of the Randal-Sundrum model {\em per se}, while in our model
the core of the wall tends to destabilize the Higgs potential.

\section*{Acknowledgments}

We thank Gregory Gabadadze, A. Losev and Mikhail Voloshin for fruitful discussions on several aspects of this
work. We also thank Elena Caceres and Nemanja Kaloper for useful communications. The work is
supported in part by the US Department of Energy under the grant number DE-FG-02-94-ER-40823.


\begin{thebibliography}{99}

\bibitem{add}
N.~Arkani-Hamed, S.~Dimopoulos and G.~R.~Dvali,
Phys.\ Lett.\ B {\bf 429}, 263 (1998)
[hep-ph/9803315].

\bibitem{ant}
I.~Antoniadis, N.~Arkani-Hamed, S.~Dimopoulos and G.~R.~Dvali,
Phys.\ Lett.\ B {\bf 436}, 257 (1998)
[hep-ph/9804398].

\bibitem{add1}
N.~Arkani-Hamed, S.~Dimopoulos and G.~R.~Dvali,
Phys.\ Rev.\ D {\bf 59}, 086004 (1999)
[hep-ph/9807344].

\bibitem{polchinski}
J.~Polchinski,
Phys.\ Rev.\ Lett.\  {\bf 75}, 4724 (1995)
[hep-th/9510017].

\bibitem{local1}
V.~A.~Rubakov and M.~E.~Shaposhnikov,
Phys.\ Lett.\ B {\bf 125}, 136 (1983).

\bibitem{local2}
G.~R.~Dvali and M.~A.~Shifman,
Phys.\ Lett.\ B {\bf 396}, 64 (1997)
[Erratum-ibid.\ B {\bf 407}, 452 (1997)]
[hep-th/9612128].

\bibitem{DR}
S.~L.~Dubovsky and V.~A.~Rubakov,
Int.\ J.\ Mod.\ Phys.\ A {\bf 16}, 4331 (2001)
[hep-th/0105243].

\bibitem{susygaugebreak}
See $e.g.$ the recent discussion:
R.~Barbieri, L.~J.~Hall and Y.~Nomura,
Phys.\ Rev.\ D {\bf 63}, 105007 (2001)
[hep-ph/0011311].

\bibitem{gene}
N.~Arkani-Hamed and M.~Schmaltz,
Phys.\ Rev.\ D {\bf 61}, 033005 (2000)
[hep-ph/9903417];
G.~R.~Dvali and M.~A.~Shifman,
Phys.\ Lett.\ B {\bf 475}, 295 (2000)
[hep-ph/0001072];
D.~E.~Kaplan and T.~M.~Tait,
JHEP {\bf 0006}, 020 (2000)
[hep-ph/0004200];
B.~A.~Dobrescu and E.~Poppitz,
Phys.\ Rev.\ Lett.\  {\bf 87}, 031801 (2001)
[hep-ph/0102010];
N.~Borghini, Y.~Gouverneur and M.~H.~Tytgat,
hep-ph/0108094.


\bibitem{neut}
G.~R.~Dvali and A.~Y.~Smirnov,
Nucl.\ Phys.\ B {\bf 563}, 63 (1999)
[hep-ph/9904211];
N.~Arkani-Hamed, S.~Dimopoulos, G.~R.~Dvali and J.~March-Russell,
hep-ph/9811448.

\bibitem{proton}
See $e.g.$
A.~Aranda and C.~D.~Carone,
Phys.\ Rev.\ D {\bf 63}, 075012 (2001)
[hep-ph/0012092].

\bibitem{conf1}
H. Weyl, {\it Space--Time--Matter} (Dover, New York, 1952);
R.~H.~Dicke,
Phys.\ Rev.\  {\bf 125}, 2163 (1962).

\bibitem{conf2}
J.~D.~Bekenstein and A.~Meisels,
Phys.\ Rev.\ D {\bf 22}, 1313 (1980).

\bibitem{rs1}
L.~J.~Randall and R.~Sundrum,
Phys.\ Rev.\ Lett.\  {\bf 83}, 3370 (1999)
[hep-ph/9905221].


\bibitem{kaloper}
N.~Kaloper, J.~March-Russell, G.~D.~Starkman and M.~Trodden,
Phys.\ Rev.\ Lett.\  {\bf 85}, 928 (2000)
[hep-ph/0002001].

\bibitem{cancel}
N.~Arkani-Hamed, S.~Dimopoulos and J.~March-Russell,
Phys.\ Rev.\ D {\bf 63}, 064020 (2001)
[hep-th/9809124].

\bibitem{exp}
C.~D.~Hoyle, U.~Schmidt, B.~R.~Heckel, E.~G.~Adelberger, J.~H.~Gundlach, D.~J.~Kapner 
and H.~E.~Swanson,
Phys.\ Rev.\ Lett.\  {\bf 86}, 1418 (2001)
[hep-ph/0011014].

\bibitem{ads}
P.~Breitenlohner and D.~Z.~Freedman,
Phys.\ Lett.\ B {\bf 115}, 197 (1982);
Annals Phys.\  {\bf 144}, 249 (1982);
L.~Mezincescu and P.~K.~Townsend,
Annals Phys.\  {\bf 160}, 406 (1985).

\bibitem{others}
 N.~ Cornish and D.~Spergel, math.DG/9906017;
M.~Trodden,
hep-th/0010032;
G.~D.~Starkman,
Class.\ Quant.\ Grav.\  {\bf 15}, 2529 (1998).

\bibitem{PS}
J.~Polchinski and M.~J.~Strassler,
Phys.\ Rev.\ Lett.\  {\bf 88}, 031601 (2002)
[hep-th/0109174].

\bibitem{DS}
G.~R.~Dvali and M.~A.~Shifman,
Nucl.\ Phys.\ B {\bf 504}, 127 (1997)
[hep-th/9611213].


\bibitem{cancelp}
R.~Sundrum,
Phys.\ Rev.\ D {\bf 59}, 085010 (1999)
[hep-ph/9807348];
Phys.\ Rev.\ D {\bf 59}, 085009 (1999)
[hep-ph/9805471].

\bibitem{cosmo}
G.~D.~Starkman, D.~Stojkovic and M.~Trodden,
Phys.\ Rev.\ Lett.\  {\bf 87}, 231303 (2001)
[hep-th/0106143];
Phys.\ Rev.\ D {\bf 63}, 103511 (2001)
[hep-th/0012226].


\end{thebibliography}
\end{document}